\documentclass[onecolumn]{IEEEtran}
\usepackage{setspace,amsmath,latexsym,cite,amssymb,epsfig,amsfonts}
\usepackage{url,cite}
\usepackage{graphicx}
\usepackage{psfrag}
\usepackage{color}
\usepackage{epstopdf}
\usepackage{multirow}
\usepackage{standalone}
\usepackage{pgfplots}
\pgfplotsset{compat=newest}
\usepackage[font={normalsize}]{caption}
\allowdisplaybreaks

\title{Energy-Efficient Heterogeneous Cellular Networks with Spectrum Underlay and Overlay Access}
\author{Jie Tang,~\emph{Member},~\emph{IEEE}, Daniel K. C. So,~\emph{Senior}~\emph{Member},~\emph{IEEE},\\ Emad Alsusa,~\emph{Senior}~\emph{Member},~\emph{IEEE}, Khairi Ashour Hamdi,~\emph{Senior}~\emph{Member},~\emph{IEEE},\\\ Arman Shojaeifard,~\emph{Member},~\emph{IEEE}, and Kai-Kit Wong,~\emph{Fellow}~\emph{IEEE}
\thanks{J. Tang is with the School of Electronic and Information Engineering, South China University of Technology, Guangzhou, China (eejtang@scut.edu.cn).\par D. K. C. So, E. Alsusa, and K. A. Hamdi are with the School of Electrical and Electronic Engineering, University of Manchester, Manchester, United Kingdom (d.so@manchester.ac.uk; e.alsusa@manchester.ac.uk; k.hamdi@manchester.ac.uk). \par A. Shojaeifard and K.-K Wong are with the Department of Electronic and Electrical Engineering, University College London, London WC1E 7JE, United Kingdom. (a.shojaeifard@ucl.ac.uk; kai-kit.wong@ucl.ac.uk). \par This work has been supported by the South China University of Technology Xinghua Talents Program under grant J2RS-D615042III and Engineering and Physical Sciences Research Council (EPSRC) under grants EP/J021768/1 and EP/N008219/1.}
}

\begin{document}

\maketitle

\begin{abstract}

In this paper, we provide joint subcarrier assignment and power allocation schemes for quality-of-service (QoS)-constrained energy-efficiency (EE) optimization in the downlink of an orthogonal frequency division multiple access (OFDMA)-based two-tier heterogeneous cellular network (HCN). Considering underlay transmission, where spectrum-efficiency (SE) is fully exploited, the EE solution involves tackling a complex mixed-combinatorial and non-convex optimization problem. With appropriate decomposition of the original problem and leveraging on the quasi-concavity of the EE function, we propose a dual-layer resource allocation approach and provide a complete solution using difference-of-two-concave-functions approximation, successive convex approximation, and gradient-search methods. On the other hand, the inherent inter-tier interference from spectrum underlay access may degrade EE particularly under dense small-cell deployment and large bandwidth utilization. We therefore develop a novel resource allocation approach based on the concepts of spectrum overlay access and resource efficiency (RE) (normalized EE-SE trade-off). Specifically, the optimization procedure is separated in this case such that the macro-cell optimal RE and corresponding bandwidth is first determined, then the EE of small-cells utilizing the remaining spectrum is maximized. Simulation results confirm the theoretical findings and demonstrate that the proposed resource allocation schemes can approach the optimal EE with each strategy being superior under certain system settings.

\end{abstract}

\begin{IEEEkeywords}

Heterogeneous cellular network (HCN), orthogonal frequency division multiple access (OFDMA), energy-efficiency (EE), spectrum-efficiency (SE), resource efficiency (RE).

\end{IEEEkeywords}

\section{Introduction}

The global mobile data traffic, thanks largely to the ever-growing use of applications on smart devices, increased by a tremendous $4$k times in a decade from 2005 to 2015 and is expected to further grow going into 2020 and beyond \cite{cisco}. It is well-understood that the conventional cellular network architecture using macro-cells only cannot possibly support demand going forward. This trend has driven the wireless industry to devise new technologies and standards for a new fifth-generation (5G) mobile network. A promising enabler for supporting user equipments (UEs) with increased density and quality-of-service (QoS) requirements is to deploy different types of base stations (BSs), thus forming what is referred to as heterogeneous cellular network (HCN) \cite{andrews}. The underlying air interface technology for HCN in the downlink is orthogonal frequency division multiple access (OFDMA) as specified in modern cellular standards \cite{Perez}.       

Meanwhile, spectrum-efficiency (SE), a measure of the total amount of information transmitted per unit bandwidth, has been used as a key performance indicator in the design and analysis of cellular networks \cite{Kang2002, Dai2003, Verdu2002,6336759,6262494}. More recently, particular focus has been placed on the SE maximization problem in OFDMA-based HCNs. In particular, although intra-cell interference is suppressed via exclusive channel assignment in OFDMA, considering the dense and irregular deployment of nodes in HCNs, there remains inter-cell interference from both intra- and inter-tier sources \cite{6918448}. As a result, novel interference management strategies for HCNs has been an active area of research \cite{Chandrasekhar2009,7070674,7478073}. In \cite{Kim2010}, a joint subcarrier and power allocation method is proposed to maximize SE considering densely deployed small-cells. In \cite{Giupponi}, a distributed resource allocation scheme using convex optimization is developed to enhance SE in a two-tier HCN. The authors in \cite{Venturino} propose a joint scheduling and power allocation scheme for maximizing the HCN sum rate. As a remedy to poor performance or/and high complexity, the authors in \cite{KSon} propose a novel distributed interference management scheme. In addition, taking into account the presence of UEs with heterogeneous QoS requirements, the authors in \cite{HaijunZhang} propose a joint subcarrier assignment and power allocation algorithm for the small-cells under an interference temperature limit to protect the macro-cell from harmful inter-tier interference. In \cite{DuyTrongNgo}, the problem of joint subcarrier assignment and power allocation is investigated in the context of downlink OFDMA-based HCNs where the sum rate of all small-cell UEs is maximized whilst protecting the minimum throughput requirements of the macro-cell UEs.

On the other hand, placing the focus solely on maximizing SE will lead to ever-rising network power consumption, which goes against global commitments for sustainable development jointly in terms of energy cost and envionmental factors. Energy-efficiency (EE), defined as the total amount of information delivered per unit energy, is widely recognized as an important measure for joint spectrum- and energy-efficient cellular network design \cite{7118688}. The EE optimization problem has attracted great interest in the context of OFDMA-based systems \cite{Xiong2011, Buzzi2012, Xiao2013, Xiong20133,7177127}, and more recently OFDMA-based HCNs \cite{GubongLim, ShengrongBu, RindranirinaRamamonjison, Wei-ShengLai}. In \cite{GubongLim}, a resource allocation scheme for maximizing EE in spectrum underlay access OFDMA-based HCNs is proposed. In \cite{ShengrongBu}, the authors have jointly considered EE enhancement and interference control for OFDMA-based HCNs where the problem has been formulated as a Stackelberg game. In \cite{RindranirinaRamamonjison}, the authors propose an energy-efficient spectrum sharing scheme among a macro-cell and multiple small-cells. In particular, the EE of the small-cells is maximized whilst preventing any potential severe interference leakage to the macro-cell UEs. The authors in \cite{Wei-ShengLai} further investigate the joint power allocation and admission control problem in OFDMA-based HCNs. Specifically, considering spectrum underlay access, a novel resource allocation approach has been proposed with the goal of admitting the maximum possible number of UEs whilst keeping interference below a certain threshold.

\subsection{Main Contributions}

Previous works in the literature \cite{GubongLim, ShengrongBu, RindranirinaRamamonjison, Wei-ShengLai} investigated EE optimization in spectrum sharing OFDMA-based HCNs with spectrum underlay access, where interference constraints are imposed on the small-cell BSs in order to protect the QoS requirements of the macro-cell UEs. The inherent severity of inter-tier interference from spectrum underlay access may nevertheless degrade EE in certain cases. In addition, in contrast to the state-of-the-art studies on EE maximization \cite{GubongLim, ShengrongBu, RindranirinaRamamonjison, Wei-ShengLai} where the impact of spectrum utilization is not considered, in this paper we directly include bandwidth usage in the analysis by modeling the dynamic circuit power consumption as a linear function of the bandwidth \cite{ShunqingZhangVTC, ShunqingZhangGLOBECOM}. As a result, the global frequency reuse strategy may lead to higher circuit power consumption and degraded EE performance. On the other hand, utilizing overlay transmission, where the allocated bandwidth for the macro-cell and the small-cells are exclusive, is considered a promising strategy when it comes to densely deployed or bandwidth-abundant HCNs. In this paper, a fundamental study of EE in the context of an OFDMA-based two-tier HCN consisting of a macro-cell and multiple small-cells is provided. We consider both underlay and overlay transmission strategies and provide joint subcarrier assignment and power allocation schemes for maximizing EE subject to satisfying QoS constraints. 

We first consider spectrum underlay access under constraints in terms of minimum throughput requirements for the macro-cell and the small-cells UEs as well as maximum transmit powers of the different BS types. The EE maximization problem under this setup, involving the apportion of limited radio resources to different UEs in different cells, is mixed-combinatorial and non-convex, and hence very challenging to solve. In order to tackle this, we decompose the original problem into a series of sub-problems with single inequality constraints. Based on the quasi-concavity of the EE function, a dual-layer resource allocation approach is accordingly proposed for solving each sub-problem. We provide a complete solution where the inner-layer is solved using difference-of-two-concave-functions approximation and successive convex approximation while gradient-search is invoked for the outer-layer. The proposed joint subcarrier assignment and power allocation algorithm using underlay transmission may not be energy-efficient in certain cellular envionments such as dense (due to severe inter-tier interference) or bandwidth-abundant (due to high circuit power) cases. In additional, the computational complexity for the proposed underay-based approach can be high when the number of subcarriers or/and the number of UEs is relatively large. Consequently, motivated by our previous work on resource efficiency (RE) \cite{TangRE} and the idea of spectrum overlay access as an effective inter-tier interference mitigation strategy, a novel transmission scheme for the two-tier HCN is proposed where the optimization procedure for the macro-cell and the small-cells is separated. Specifically, we first optimize the RE at the macro-cell and determine the corresponding optimal bandwidth using joint subcarrier assignment and power allocation. We then assign the remaining spectrum to the small-cells and accordingly optimize their EE. Numerical results validate the effectiveness of the proposed algorithms and show that each algorithm has its strong point depending on scenario parameters such as the small-cells density and the bandwidth utilization.

\subsection{Organization}

The remainder of this paper is organized as follows. The system model and problem formulation is given in Section II. In Section III, the fundamentals for EE optimization in OFDMA-based HCNs is studied, where in particular the EE optimization problem with multiple inequality constraints is transformed into a multiple single inequality problems. In Section IV, joint subcarrier assignment and power allocation algorithm for the inner-layer is first introduced by exploiting some properties of the optimization problem. In particular, an efficient solution based on difference-of-two-concave-functions approximation for the inner-layer process is developed, followed by a complete solution to the dual-layer scheme. In Section V, a novel low-complexity algorithm is developed based on overlay transmission. Simulation results are provided in Section VI and conclusions are drawn in Section VII.

\section{Preliminaries}

In this section, we describe the two-tier OFDMA-based HCN setup under consideration. The QoS-constrained EE optimization problem is then mathematically formulated.

\subsection{System Model}

We consider the downlink of an OFDMA-based two-tier HCN comprising of a macro-cell and $L$ small-cells. The set of cells is denoted using $\mathcal{L} = \{0, 1, 2, \cdots, L\}$, where indexes $0$ and $\{1, 2, \cdots, L\}$ correspond to the macro-cell and the small-cells, respectively. In addition, we assume that there are ${K}_0$ macro-cell UEs (MUEs) and $K_l$ small-cell UEs (SUEs). For simplicity, the index of the UEs (MUEs and SUEs) associated with cell $l \in \mathcal{L}$ is denoted with $\mathcal{K}_l$. The HCN total available spectrum, $W_{tot}$, is divided into $N_{tot}$ subcarriers with each having a bandwidth of $W_C = \frac{W_{tot}}{N_{tot}}$. Specifically, the set of all accessible frequencies is denoted with $\mathcal{N}$ (where $|\mathcal{N}|$ = $N_{tot}$). We consider exclusive channel assignment, where any OFDMA subcarrier can only be employed by at most one UE in a given cell at a given time, in order to prevent harmful collision. Note that the UE-BS association is considered fixed during runtime. 

The channel power gain from the cell-$m$ BS to the cell-$l$ $k$-th UE over subcarrier $n$ is denoted with $h_{[k,l,m]}^{n}$. The received signal-to-interference-plus-noise ratio (SINR) at the cell-$l$ $k$-th UE over subcarrier $n$ can be formulated as \cite{7277029}
\begin{equation}
{\label{received_SINR} \gamma_{[k,l]}^n = \frac{h_{[k,l,l]}^{n} p_{l}^{n}}{\sum_{m \in \mathcal{L}\backslash\{l\}}  h_{[k,l,m]}^{n} p_{m}^{n} +\sigma^{n}_{[k,l]}}}
\end{equation}
where $p_{l}^{n}$ is the allocated transmit power of the cell-$l$ BS over subcarrier $n$ and $\sigma^{n}_{[k,l]}$ is the noise power at the cell-$l$ $k$-th UE over subcarrier $n$. Note that intra-cell interference is avoided under exclusive channel assignment. We thus can express the rate for the cell-$l$ $k$-th UE over subcarrier $n$ using 
\begin{equation}\label{data rate}
{r_{[k,l]}^n = W_C\log_2 \left( 1+ \gamma_{[k,l]}^n \right)}.
\end{equation}

Hence, the total throughput in cell-$l$ is given by
\begin{equation}\label{throughput_cell}
{C_{l} = \sum_{n \in \mathcal{N}} \sum_{k \in \mathcal{K}_l} \rho_{[k,l]}^{n} r_{[k,l]}^n}
\end{equation}
where $\rho_{[k,l]}^{n} \in \{1, 0\}$ indicates whether or not the $n^{th}$ subcarrier is assigned to the UE $[k,l]$. Considering that BSs are the dominant sources of energy consumption in cellular networks, we approximate the HCN overall power consumption using the following linear power model \cite{Cui2005} 
\begin{equation}\label{overall power_ge}
P = \zeta P_T + P_C
\end{equation}
where $\zeta$, $P_T$ and $P_C$ denote the BS reciprocal of drain efficiency of the power amplifier, transmission power, and circuit power consumption, respectively. Furthermore, the circuit energy consumption consists of static (fixed) and dynamic parts, where the latter depends on the active links parameters. Motivated by the approach in \cite{ShunqingZhangVTC} and \cite{ShunqingZhangGLOBECOM}, the circuit power consumption is considered to be proportional to the total utilized bandwidth for transmission. Consequently, the total circuit power can be written as
\begin{equation}\label{circuit_power}
P_c = P_s + \gamma W
\end{equation}
where $P_s$ is the static circuit power in transmission mode and $\gamma$ is a constant corresponding to the dynamic power consumption per unit bandwidth. As a result, the total power consumption in the two-tier OFDMA-based HCN under consideration is defined as
\begin{equation}\label{overall power}
{P = \sum_{l \in \mathcal{L}}(\zeta P_T^{[l]} + P_C^{[l]})}
\end{equation}
where $P_T^{[l]}$ and $P_C^{[l]}$ are the transmission power and the circuit power at the cell-$l$ BS.

\subsection{Problem Formulation}

Recall that EE is defined as the total number of successfully delivered bits per unit energy. The two-tier OFDMA-based HCN EE in the downlink can hence be described using the following equation
\begin{equation}\label{EE}
\lambda_{EE} \triangleq \frac{C}{P} = \frac{\sum_{l \in \mathcal{L}}C_{l}}{\sum_{l \in \mathcal{L}}(\zeta P_T^{[l]} + P_C^{[l]}) }
\end{equation}
where $C$ and $P$ are respectively used to denote the total data rate and the total power consumption of the HCN. Here, we are concerned with the problem of achieving high EE whilst guaranteeing the required QoS constraints in each cell under limited bandwidth and transmit power resources. Hence, the we formulate an optimization problem for maximizing EE under a series of (minimum) throughput requirements and maximum power budgets. Accordingly, we formulate the EE optimization problem for the two-tier OFDMA-based HCN as
\begin{align}
& \max_{\rho_{[k,l]}^{n}, p_{l}^{n}} ~~ \lambda_{EE} \label{objective}\\
& \textrm{s.t.} ~~ \sum_{n \in \mathcal{N}} {p}_{l}^n\leq P_{max}^{[l]}, ~ \forall l \in \mathcal{L}, \label{constraint1}\\
& ~~~~~~ \sum_{n \in \mathcal{N}} \sum_{k \in \mathcal{K}_l} \rho_{[k,l]}^{n}r_{[k,l]}^n \geq \delta_{small}, ~\forall l \in \mathcal{L}\backslash\{0\}, \label{constraint2}\\
& ~~~~~~ \sum_{n \in \mathcal{N}} \sum_{k \in \mathcal{K}_0} \rho_{[k,0]}^{n}r_{[k,0]}^n \geq \delta_{macro}, \label{constraint3}\\
& ~~~~~~ \sum_{k \in \mathcal{K}_l} \rho_{[k,l]}^{n} = 1, \forall n \in \mathcal{N}, \forall l \in \mathcal{L}, \label{constraint4}
\end{align}
where $P_{max}^{[l]}$ is the maximum transmit power of the cell-$l$ BS and $\delta_{macro}$ ($\delta_{small}$) correspond to the UEs minimum throughput requirements in the macro-cell (small-cells), respectively. Therefore, constraints (\ref{constraint1})-(\ref{constraint3}) are used to guarantee the maximum power budget and the minimum throughput target in each cell. In addition, the constraint in (\ref{constraint4}) corresponds to the exclusive subcarrier assignment strategy in any cell. 

The EE optimization problem here, which considers join subcarrier assignment and power allocation in the presence of inter-cell interference, is mixed-combinatorial and non-convex. The solution is therefore nontrivial and cannot be obtained directly. As a result, in the following sections, we develop two different resource allocation approaches considering both spectrum underlay and overlay access. 

\section{Fundamentals of EE Optimization in OFDMA-based HCNs with Spectrum Underlay Access}

With spectrum underlay access, the small-cells share the available radio spectrum with the macro-cell and hence introduce inter-tier interference which renders the resource allocation problem significantly more challenging to tackle. In addition, the variables for subcarrier assignment and power allocation are are coupled together and hence the non-convex optimization problem in (\ref{objective})-(\ref{constraint4}) is extremely difficult to solve. In this section, we provide a fundamental study for energy-efficient design in OFDMA-based underlay HCNs. In particular, a relationship between optimal EE and achievable throughput is derived as in the following theorem.

\textbf{\emph{Theorem I.}} \emph{For any rate vector for the macro-cell and the small-cells that satisfies the minimum throughput constraint, $\textbf{C} \geq \boldsymbol{\delta}$, achieved with subcarrier assignment ${\rho}_{[k,l]}^n$ and power allocation ${p}_{l}^n, \forall~(l,n) \in (\mathcal{L}, \mathcal{N})$, the maximum achievable EE, namely,}
\begin{align}
& {\lambda}^*_{EE}(\textbf{C}) \triangleq \max_{{\rho}_{[k,l]}^n,{p}_{l}^n}~\lambda_{EE} \label{objectivepo}\\
& \textrm{s.t.} ~~\sum_{n \in \mathcal{N}} {p}_{l}^n\leq P_{max}^{[l]},~\forall~ l \in \mathcal{L}, \label{constraint1po}\\
& ~~~~~~ \textbf{C} \geq \boldsymbol{\delta}, ~\forall~ l \in \mathcal{L}, \label{constraint22po}\\
& ~~~~~ \sum_{k \in \mathcal{K}_l} \rho_{[k,l]}^{n} = 1, \forall~n \in \mathcal{N}, \forall~l \in \mathcal{L}, \label{constraint2po}
\end{align}
\emph{where $\textbf{C} = [C_0~C_1~\cdots~C_L]$ and $\boldsymbol{\delta} = [\delta_{macro}~\delta_{small}~\cdots~\delta_{small}]$, is strictly quasi-concave in $\textbf{C}$.}
\emph{Proof:} See Appendix A.

By definition, a continuous and strictly quasi-concave function has a unique maximum value over a finite domain \cite{Boyd04}. Therefore, \emph{Theorem I} indicates that there always exists a unique EE solution. On the other hand, the original EE optimization problem is very challenging to solve due to the multiple inequality constraints in (\ref{constraint2})-(\ref{constraint3}). By extending the Lagrange dual decomposition method for single-cell multi-carrier systems \cite{Xiong2011} to our OFDMA-based underlay HCN setup, the gradient ascent approach can be invoked to generate $\textbf{C}, l =0,1,2,\cdots$, and
\begin{equation}\label{LDD}
{ \textbf{C}(n+1) = [\textbf{C}(n) + \mu \nabla {\lambda}_{EE}(\textbf{C}(n))]^+.} 
\end{equation}
However, a closed-form expression for the gradient of this vector does not exist, hence, it is impossible to directly employ (\ref{LDD}) in order to obtain the solution in (\ref{objective})-(\ref{constraint4}). However, we can transform the gradient ascent method in (\ref{LDD}) using the following approach
\begin{gather}
C_{0}(n+1) = [{C}_{0}(n) + \mu \nabla {\lambda}_{EE}({C}_0(n))]^+, \nonumber \\
\vdots \nonumber \\
C_{L}(n+1) = [{C}_{L}(n) + \mu \nabla {\lambda}_{EE}({C}_{L}(n))]^+. \label{LDD3}
\end{gather}
As a result, the vector gradient can be alternatively decomposed into multiple scalar gradient, thus making the optimization problem relatively easier to solve. Accordingly, we propose an iterative resource allocation scheme to tackle the EE optimization problem for two-tier OFDMA-based HCNs. Similar to the gradient decomposition approach, by keeping one minimum throughput/rate constraint at a time and setting all others as equality constraints (fixed throughput/rate), the optimization problem with multiple inequality constraints can be decomposed into a series of optimization problems with single inequality constraint. Specifically, under inequality constraint (minimum throughput/rate requirement) for cell-$l^*$, all other cells will be under equality constraints. Therefore, the EE maximization problem is transformed into
\begin{align}
& \max_{\rho_{[k,l]}^{n}, p_{l}^{n}}~~ \lambda_{EE} \label{objectivepp}\\
& ~~\textrm{s.t.}~ {C}_l = \bar{{C}}_l, ~\forall~ l \in \mathcal{L}\backslash\{l^*\}, \label{constraint2pp}\\
& ~~~~~ \sum_{n \in \mathcal{N}} \sum_{k \in \mathcal{K}_l^*} \rho_{[k,l^*]}^{n}{r}_{[k,l^*]}^{n} \geq {\delta}_l^*, \label{constraint3pp}\\
& ~~~~~ \sum_{k \in \mathcal{K}_l} \rho_{[k,l]}^{n} = 1, \forall~n \in \mathcal{N}, \forall~l \in \mathcal{L}, \label{constraint4pp}
\end{align}
where $\bar{C}_{l}$ represents the optimal throughput (for all other cells apart from cell $l^*$) obtained from the previous iteration. Under this setup, we can obtain the solution for cell $l^*$, $\bar{C}_{l^*}$ by solving the above single inequality constrained optimization problem. Once we obtain the updated throughput $\bar{C}_{l^*}$, the next cell is placed under inequality constraint (minimum throughput requirement), i.e., $\forall~l^*+1 \in \mathcal{L}$, while all other cells have updated throughput values from the previous iteration. In particular, we can rewrite the constraints in (\ref{constraint2pp})-(\ref{constraint3pp}) as
\begin{gather}
{C}_l = \bar{{C}}_l, ~\forall~ l \in \mathcal{L}\backslash\{l^*+1\}, \label{constraint3ppp} \\ 
\sum_{n \in \mathcal{N}} \sum_{k \in \mathcal{K}_{l^*+1}} \rho_{[k,l^*+1]}^{n}{r}_{[k,{l^*+1}]}^{n} \geq {\delta}_{l^*+1}. \label{constraint4ppp}
\end{gather}
The current maximum EE value is stored in the buffer and the corresponding optimal rate for cell $l^*+1$ $\bar{C}_{l^*+1}$ is updated. This process is repeated for all cells until convergence, i.e., $\lambda^{opt}_{EE}(n+1) - \lambda^{opt}_{EE}(n) \leq \varepsilon$. We provide a pseudocode for the proposed iterative resource allocation scheme:
\begin{itemize}
\item[(1)] Initialize $l^*$ as the first cell in $\mathcal{L}$ with inequality constraint;
\item[(2)] Tackle the problem in (\ref{objectivepp})-(\ref{constraint4pp}) and store $\lambda^{opt}_{EE}(n)$ in the buffer;
\item[(3)] Modify the constraints using (\ref{constraint3ppp})-(\ref{constraint4ppp}) and update the corresponding rateS;
\item[(4)] Repeat steps (2) and (3) until convergence $\lambda^{opt}_{EE}(n+1) - \lambda^{opt}_{EE}(n) \leq \varepsilon$.
\end{itemize}
The decomposed EE optimization problem in (\ref{objectivepp})-(\ref{constraint4pp}) has a single inequality constraint. With a fundamental study of the problem, we can arrive at the following theorem.

\textbf{\emph{Theorem II.}} \emph{The maximum EE achieved with a minimum throughput for cell-$l^*$, $C_{l^*} \geq \delta_{l^*}$, subcarrier assignment ${\rho}_{[k,l]}^n$, and power allocation ${p}_{l}^n, \forall(l,n)\in(\mathcal{L},\mathcal{N})$, namely,}
\begin{align}
& {\lambda}^*_{EE}(C_{l^*}) \triangleq \max_{\rho_{[k,l]}^{n}, p_{l}^{n}}~\lambda_{EE} \label{objectivep}\\
& \textrm{s.t.} ~{C}_l = \bar{{C}}_l, ~\forall~ l \in \mathcal{L}\backslash\{l^*\}, \label{constraint1p}\\
& \sum_{n \in \mathcal{N}} \sum_{k \in \mathcal{K}_l^*} \rho_{[k,l^*]}^{n}{r}_{[k,l^*]}^{n} = C_{l^*} \geq {\delta}_l^*, \label{constraint2p}\\
& \sum_{k \in \mathcal{K}_l} \rho_{[k,l]}^{n} = 1, \forall~n \in \mathcal{N}, \forall~l \in \mathcal{L} \label{constraint3p}
\end{align}
\emph{is strictly quasi-concave in $C_{l^*}$.}\\
\emph{Proof:} \emph{Theorem II} is a special case of \emph{Theorem I}, thus a similar proof to that in Appendix A can be applied here.

The function quasi-concavity property guarantees the existence of a unique maximum, hence \emph{Theorem II} proves the existence of a unique EE solution. Moreover, the quasi-concavity of EE optimization problem further indicates that $ \lambda_{EE}(C_{l^*})$ either decreases or first increases and then decreases with $C_{l^*}$
Thus, problem (\ref{objectivep})-(\ref{constraint3p}) can be solved through a dual-layer decomposition method using the processes:
\begin{itemize}
\item[(i)] Inner-layer: Finds the maximum EE in cell-$l^*$, $\lambda^*_{EE}(C_{l^*})$, under a fixed rate, $C_{l^*}$.
\item[(ii)] Outer-layer: Obtains the optimal EE, $\lambda^{opt}_{EE}$, using heuristic search.
\end{itemize}
Note that the key challenge for adopting the proposed dual-layer decomposition method lies in the inner-layer mechanism, as discussed in the following section.

\section{Joint Subcarrier Assignment and Power Allocation for EE Maximization}

In this section, we provide a joint subcarrier assignment and power allocation method for the inner-layer by exploiting the fundamental properties of the optimization problem. A complete solution to the proposed dual-layer resource allocation approach is then presented.

Given that the optimization problem in (\ref{objectivep})-(\ref{constraint3p}) involves fixed throughput requirements (equality constraints), it can be equivalently expressed in terms of the the following power minimization problem
\begin{align}
& \min_{{\rho}_{[k,l]}^n,{p}_{l}^n} \sum_{l \in \mathcal{L}}\sum_{n \in \mathcal{N}} {p}_{l}^n \label{objectivepm}\\
& \textrm{s.t.}~~ \sum_{n \in \mathcal{N}} \sum_{k \in \mathcal{K}_l} \rho_{[k,l]}^{n}{r}_{[k,l]}^{n} = C_{l},~\forall~ l \in \mathcal{L}, \label{constraint1pm}\\
& ~~~~~ \sum_{k \in \mathcal{K}_l} \rho_{[k,l]}^{n} = 1, \forall~n \in \mathcal{N}, \forall~l \in \mathcal{L}, \label{constraint3pm}\\
& ~~~~~ \sum_{n \in \mathcal{N}} {p}_{l}^n\leq P_{max}^{[l]},~\forall~ l \in \mathcal{L}. \label{constraint4pm}
\end{align}
The above power minimization problem involves subcarrier assignment and power allocation, therefore, we can extend the iterative approach proposed in \cite{Venturino} to a HCN scenario. 

The joint subcarrier assignment and power allocation process can be separated as
\begin{equation}\label{iter_algo}
{\underbrace{\boldsymbol{\rho}[0] \rightarrow \textbf{p}[0]}_{\textmd{Initialization}}\rightarrow \cdots \underbrace{\boldsymbol{\rho}[t] \rightarrow \textbf{p}[t]}_{\textmd{Iteration t}}\rightarrow \underbrace{\boldsymbol{\rho}^{opt} \rightarrow \textbf{p}^{opt}}_{\textmd{Optimal Solution}}.}
\end{equation}
where $\textbf{p}^n = [p^n_0, p^n_1,\cdots, p^n_L]$, $\textbf{p}_l = [p^1_l, p^2_l,\cdots, p^N_l]$, $\textbf{p} = \textmd{vec}[\textbf{p}_0,\textbf{p}_1,\cdots,\textbf{p}_L]$, $\boldsymbol{\rho}_{[k,l]} = [\rho_{[k,l]}^1, \rho_{[k,l]}^2,$ $\cdots,\rho_{[k,l]}^N]$, $\boldsymbol{\rho}_{l}=\textmd{vec}[\rho_{[1,l]},\rho_{[2,l]},\cdots,\rho_{[K_l,l]}]$ and $\boldsymbol{\rho} = \textmd{vec}[\rho_{[0]},\rho_{[1]},\cdots,\rho_{[L]}]$. Note that the number inside the square bracket denotes the iteration number. Next, we evaluate a feasible solution $(\boldsymbol{\rho}[0],\textbf{p}[0])$. At the initial moment of each iteration $t$, based on a given power allocation $\textbf{p}[t-1]$ from the last iteration, we solve the subcarrier assignment problem and obtain the optimal $\boldsymbol{\rho}[t]$. We then find the optimal power allocation $\textbf{p}[t]$ based on the fixed $\boldsymbol{\rho}[t]$ obtained from the previous step. This process is repeated until convergence, i.e., no further EE improvement is realized. Therefore, this iterative resource allocation approach simplifies the original EE problem by separating it into two sub-problems, namely the subcarrier assignment process and the power allocation process. More importantly, the number of variables is decreased by nearly a half in each sub-problem hence allowing for more tractable algorithm designs.

\subsection{Optimal Subcarrier Assignment for Power Minimization Problem}

Having a fixed power allocation $\textbf{p}[t-1]$ obtained from the last iteration, we attempt to obtain the optimal subcarrier assignment $\boldsymbol{\rho}[t]$ at iteration $t$. The optimization problem in (\ref{objectivepm})-(\ref{constraint4pm}) is accordingly converted to
\begin{align}
& \max_{{\rho}_{[k,l]}^n} \sum_{l \in \mathcal{L}} \sum_{n \in \mathcal{N}} \sum_{k \in \mathcal{K}_l} {r}_{[k,l]}^{n} ({\rho}_{[k,l]}^n, \textbf{p}^n[t-1]) \label{objectivepms} \\
& ~~ \sum_{k \in \mathcal{K}_l} \rho_{[k,l]}^{n} = 1, \forall~n \in \mathcal{N}, \forall~l \in \mathcal{L} \label{constraint3pms}
\end{align}
where ${r}_{[k,l]}^{n} ({\rho}_{[k,l]}^n, \textbf{p}^n[t-1])$ denotes the rate function with respect to the subcarrier assignment ${\rho}_{[k,l]}^n$ and the power allocation result obtained from the previous iteration. We can therefore arrive at the following theorem.

\textbf{\emph{Theorem III.}} \emph{The solution of (\ref{objectivepms})-(\ref{constraint3pms}) involves assigning subcarriers to UEs that consume the lowest power (i.e., the highest SINR) on those subcarriers.}\\
\emph{Proof:} see Appendix B.

The optimal subcarrier assignment for all UEs (SUEs and MUEs) can be found using \emph{Theorem III}, thus avoiding the need for an exhaustive search approach which here would be exponentially computationally complex in the number of subcarriers. Furthermore, by extending the approach in \cite{Ngo} to a HCN scenario, the subcarrier assignment problem can be decomposed into $L$ sub-problems such that
\begin{align}
& \max_{{\rho}_{[k,l]}^n} \sum_{n \in \mathcal{N}} \sum_{k \in \mathcal{K}_l} {r}_{[k,l]}^{n} ({\rho}_{[k,l]}^n, \textbf{p}^n[t-1]) \label{objectivepmsd}\\
& \textrm{s.t.}~~ \sum_{n \in \mathcal{N}} \sum_{k \in \mathcal{K}_l} \rho_{[k,l]}^{n} = 1, \forall~n \in \mathcal{N}. \label{constraint1pmsd}
\end{align}
Recall from \emph{Theorem III} that the optimal solution of (\ref{objectivepmsd})-(\ref{constraint1pmsd}) is to assign each subcarrier to the UE with the highest SINR. We can therefore conclude the optimal subcarrier assignment strategy for cell $l \in \mathcal{L}$ at iteration $t$ using
\begin{equation}\label{rho_alocation}
\rho_{[k,l]}^{n}[t]~=~\rho_{[k,l]}^{n*}~=~
\begin{cases}
1,~\textmd{if}~k = \textmd{arg}~\textmd{max}_{k \in \mathcal{K}_l}{r}_{[k,l]}^{n} (\textbf{p}^n[t-1]) \\
0,~\textmd{otherwise}
\end{cases}.
\end{equation}

\subsection{Optimal Power Allocation for Power Minimization Problem}

After determining the optimal subcarrier assignment $\boldsymbol{\rho}[t]$ at iteration $t$, we aim to find the optimal power allocation. Therefore, problem (\ref{objectivepm})-(\ref{constraint4pm}) is now converted to
\begin{align}
& \min_{{p}_{l}^n} \sum_{l \in \mathcal{L}} \sum_{n \in \mathcal{N}} {p}_{l}^n \label{objectivepmp}\\
& \textrm{s.t.}~~ \sum_{n \in \mathcal{N}} \sum_{k \in \mathcal{K}_l}\rho_{[k,l]}^{n} {r}_{[k,l]}^{n} = C_{l},~\forall~ l \in \mathcal{L}, \label{constraint1pmp}\\
& ~~~~~ \sum_{n \in \mathcal{N}} {p}_{l}^n\leq P_{max}^{[l]},~\forall~ l \in \mathcal{L}. \label{constraint3pmp} 
\end{align}
It is easy to note that the power allocation problem (\ref{objectivepmp})-(\ref{constraint3pmp}) is non-convex as a result of the non-convexity of the SINR and corresponding rate functions. Hence, we extend the successive convex approximation approach proposed in \cite{Marks78} to our work in order to solve the above non-convex problem. The methodology can be described as:
\begin{itemize}
\item[1] Initialize a power vector $\textbf{p}[0]$ and $t_p = 1$. 
\item[2] Create the $t_p$-th convex sub-problem by estimating the non-concave rate function (involving SINR) with some concave function based on the previous result $\textbf{p}[t_p-1]$.
\item[3] Tackle the $t_p$-th sub-problem to achieve the solution $\textbf{p}[t_p]$, and accordingly update the approximation parameters in Step 2.
\item[4] Update $t_p = t_p+1$ and iterate this process until $\textbf{p}[t_p]$ converges.
\end{itemize}
In the following part, an efficient power allocation approach based on difference-of-two-concave-functions approximation is proposed to update Step 2 and 3.

To solve Step 2 and 3 in the above successive convex approximation process, we formulate the data rate function (\ref{data rate}) in a difference-of-two-concave-functions approximation form 
\begin{equation}\label{data rate_app}
{\sum_{n \in \mathcal{N}} r_{[k,l]}^n (\textbf{p}^n) = f_l(\textbf{p}) - g_l(\textbf{p}),}
\end{equation}
where $f_l(\textbf{p})$ and $g_l(\textbf{p})$ are respectively representing two concave functions which are defined as
\begin{equation}
{\label{fl}   f_l(\textbf{p}) =  \sum_{n \in \mathcal{N}} \ln (\sum_{m \in \mathcal{L}}  h_{[k,l,m]}^{n} p_{m}^{n} +\sigma^{n}_{[k,l]}  )  }
\end{equation}
and
\begin{equation}
{\label{gl}   g_l(\textbf{p}) =  \sum_{n \in \mathcal{N}} \ln (\sum_{m \in \mathcal{L}\backslash\{l\}}  h_{[k,l,m]}^{n} p_{m}^{n} +\sigma^{n}_{[k,l]}  )  }. 
\end{equation}
We then approximate $g_l(\textbf{p})$ based on a fixed $\textbf{p}[t_p-1]$ (obtain from iteration $t_p-1$) such that \cite{Kha12}
\begin{equation}
{\label{gl_app}  g_l(\textbf{p}) \approx  g_l(\textbf{p}[t_p - 1]) + \nabla g_l^T[t_p - 1](\textbf{p} - \textbf{p}[t_p - 1]) } 
\end{equation}
where $\nabla g_l(\textbf{p})$ is a vector with length $(L + 1)N$, and its corresponding entry is defined as
\begin{equation}\label{entry}
\nabla g_l(\textbf{p})^{(Nj+n)}~=~
\begin{cases}
0,~\textmd{if}~j = l \\
\frac{h_{[k,l,j]}^{n} }{\sum_{s \in \mathcal{L}\backslash\{l\}}  h_{[k,l,s]}^{n} p_{s}^{n} +\sigma^{n}_{[k,l]}},~\textmd{if}~j = \mathcal{L}\backslash\{l\}
\end{cases}.
\end{equation}
\noindent Therefore, combining (\ref{data rate_app}) with (\ref{entry}), we can obtain
\begin{equation}
{\label{last}  \sum_{n \in \mathcal{N}} r_{[k,l]}^n (\textbf{p}^n) \approx  f_l(\textbf{p}) - g_l(\textbf{p}[t_p - 1])- \nabla g_l^T[t_p - 1](\textbf{p} - \textbf{p}[t_p - 1]).}
\end{equation}
It should be noted that the right-hand side of the above equation is concave in \textbf{p}.

Therefore, (\ref{last}) enables us to reformulate the optimization problem (\ref{objectivepmp})-(\ref{constraint3pmp}) into a series of convex optimization sub-problems. Particularly, the $t_p$-th iteration ($t_p$ sub-problem) is established as
\begin{align}
& \min_{{p}_{l}^n} ~\| \textbf{p} \| \label{objectivepmpdc}\\
& \textrm{s.t.}~~ f_l(\textbf{p}) - g_l(\textbf{p}[t_p - 1])- \nabla g_l^T[t_p - 1](\textbf{p} - \textbf{p}[t_p - 1]) \geq C_{l},~\forall~ l \in \mathcal{L}, \label{constraint1pmpdc}\\
& ~~~~~ \sum_{n \in \mathcal{N}} {p}_{l}^n\leq P_{max}^{[l]},~\forall~ l \in \mathcal{L}, \label{constraint3pmpdc}  
\end{align}
where $\textbf{p}[t_p-1]$ has been determined from the last iteration $t_p-1$. Since the objective function and the constraints are all convex, this problem ($t_p$ sub-problem) can be efficiently solved using Branch and Bound method \cite{Boyd}. Once the sub-problem (\ref{objectivepmpdc})-(\ref{constraint3pmpdc}) is found, $\textbf{p}[t_p]$ is obtained and updated in (\ref{gl_app}) to solve the $(t_p+1)$ sub-problem in the next iteration.

\subsection{A Complete Solution to the Dual-Layer Approach}

The inner-layer, which under a fixed rate in cell-$l^*$, $C_{l^*}$, finds the maximum EE, $\lambda^*_{EE}(C_{l^*})$, can be efficiency solved based on the proposed joint subcarrier assignment and power allocation algorithm. Next, we propose an approach for the outer-layer process using gradient-search. 

With an initial setting $C_{l^*}(1)$, ${\lambda}^*_{EE}(C_{l^*}(1))$ can be obtained using the proposed joint subcarrier assignment and power allocation algorithm. On the basis of \emph{Theorem II}, we can then update $C_{l^*}$ using the following approach
\begin{equation}\label{up}
C_{l^*}(n+1)=
\begin{cases}
\frac{C_{l^*}(n)}{\varpi} & \frac{d\lambda^*_{EE}(C_{l^*})}{dC_{l^*}}{\bigg|_{C_{l^*}}(n)}<0 \\
\varpi C_{l^*}(n) & \textmd{otherwise}
\end{cases}
\end{equation}
where $\varpi > 1$ denotes the search step size. Furthermore, we need to reduce the step size $\varpi$ if the gradient $\frac{d\lambda^*_{EE}(C_{l^*})}{dC_{l^*}}$ changes its sign as
\begin{equation}\label{beta_update}
{\varpi(n+1) = \frac{\varpi(n)}{2},}
\end{equation}
and (\ref{up}) is repeated until convergence, i.e., $|{\lambda}^*_{EE}[C_{l^*}(n+1)] - {\lambda}^*_{EE}[C_{l^*}(n)]| \leq \epsilon$. The complete solution to the EE optimization problem in (\ref{objectivep})-(\ref{constraint3p}) for the OFDMA-based two-tier HCN with spectrum underlay access is summarized in Table I.
\begin{table}
\centering
\caption{A complete solution to the EE optimization problem.} 
\renewcommand{\arraystretch}{1}  
\begin{tabular} {|l|}
\hline
1)~Initialize $C_{l^*}(1) \in [\delta_{l^*}, \delta_{{l^*},(max)}]$, and set $n = 1$;\\
2)~\textbf{REPEAT}\\
3)~~~Obtain the maximum EE $\lambda^*_{EE}(C_{l^*})$ using the proposed joint subcarrier assignment \\ and power allocation approach in Section IV.A and Section IV.B;\\
4)~~~Update $C_{l^*}(n)$ using (\ref{up}); $n = n + 1$;\\
5)~\textbf{UNTIL} $|{\lambda}^*_{EE}[C_{l^*}(n+1)] - {\lambda}^*_{EE}[C_{l^*}(n)]| \leq \epsilon$;\\ \hline
\end{tabular} 
\end{table}

\section{Solution based on Spectrum Overlay Access and Resource Efficiency}

The inter-tier interference from the proposed underlay-based approach may degrade EE especially under high throughput requirements in densely deployed scenarios. In addition, considering that the available bandwidth is fully exploited by the macro-cell and the small-cells at the same time, a higher circuit power consumption and hence reduced EE performance may be incurred. Furthermore, although the proposed iterative joint subcarrier and power allocation algorithm is numerically stable, its computational complexity depends on the number of optimizing variables, which can be large if the number of subcarriers or the number of UEs is large. Hence the complexity of this scheme is comparatively high. As a result, based on the idea of spectrum overlay access and resource efficiency, we next develop a low-complexity resource allocation approach for the two-tier HCN under consideration.

\subsection{Resource Efficiency Optimization for Marco-cell}

In \cite{TangRE}, RE is defined as a weighted EE-SE trade-off using a normalizing factor $\beta$ 
\begin{equation}\label{RE}
{\lambda_{RE} \triangleq \frac{R}{P}(1+\beta\frac{\eta_P}{\eta_W})}
\end{equation}
where $\eta_P$ and $\eta_W$ respectively denote the power utilization and bandwidth utilization such that
\begin{equation}\label{power bandwidth efficiency}
{\eta_P \triangleq \frac{P}{P_{tot}}, ~~~~\eta_W \triangleq \frac{W}{W_{tot}}.}
\end{equation}
The notion behind RE maximization requires EE and SE as input vectors in a multi-objective optimization problem. Hence, there does not exist a-priori correspondence between a weight vector and a solution vector. This implies that the weights that control EE and SE has to be decided by the decision maker. In addition, it has been shown in \cite{TangRE} that the corresponding EE is decreasing with increasing $\beta$ while the corresponding SE is increasing with increasing $\beta$. As a result, we modify the RE formulation to a more generalized expression
\begin{align}
\lambda_{RE} & \triangleq \alpha \frac{R}{P} + (1-\alpha) \tau \frac{R}{P} \nonumber\\
& = \frac{R}{P} \left( \alpha +(1-\alpha) \frac{\eta_P}{\eta_W} \right) \label{RE_new2}
\end{align}
where $0 \leq \alpha \leq 1$ and $\tau = \frac{W_{tot}}{P_{tot}}$. 

The generalized RE optimization problem in the downlink of the macro-cell can be mathematically formulated as
\begin{align}
& \max_{\boldsymbol{\rho},~\textbf{p},~\alpha} \frac{ \sum_{k \in \mathcal{K}_0}\sum_{n \in \mathcal{N}}{\rho}_{[k,0]}^{n} {r}_{[k,0]}^{n} }{\zeta P_T + P_C }(\alpha+(1-\alpha)\frac{\eta_P}{\eta_W}) \label{objectivepmre}\\
& \textrm{s.t.}~~ \sum_{n \in \mathcal{N}} \sum_{k \in \mathcal{K}_0} {r}_{[k,0]}^{n} \geq \delta_{0}, \label{constraint1pmre}\\
& ~ \sum_{k \in \mathcal{K}_0} \rho_{[k,0]}^{n} = 1, \forall~n \in \mathcal{N}, \label{constraint3pmre}\\ 
& ~\sum_{n \in \mathcal{N}}\sum_{k \in \mathcal{K}_0}\rho_{[k,0]}^{n} \leq N_{tot}, \label{constraint4pmre}\\
& ~ \sum_{n \in \mathcal{N}} {p}_{0}^n\leq P_{max}^{[0]} \label{constraint5pmre}
\end{align}
where $P_T = \sum_{n \in \mathcal{N}} {p}_{0}^n$ and $P_C = P_s + \gamma W_C\sum_{k \in \mathcal{K}}\sum_{n \in \mathcal{N}}\rho_{[k,0]}^{n}$. Problem (\ref{objectivepmre})-(\ref{constraint5pmre}) is mixed-combinational and non-convex. In order to tackle this, the subcarrier assignment and power allocation procedures are separated. Specifically, we first analyze the fundamental properties of the case with a given weight $\alpha$ and a given subcarrier assignment set. The findings are summarized in the following theorem. Note that for simplicity, here, we remove the index from the macro-cell parameters, e.g., MUEs set is changed from $\mathcal{K}_0$ to $\mathcal{K}$.

\textbf{\emph{Theorem IV.}} \emph{Considering a given weight $\alpha$, subcarrier allocation vector ${\boldsymbol{\rho}}$ and its corresponding UEs set $\mathcal{S}_k (\forall k \in \mathcal{K})$, the maximum RE at a certain transmit power, $P_T$, namely,}
\begin{equation}\label{Poptimal RE}
{ {\lambda}_{RE}(P_T) \triangleq  \max_{p_{k,n}\geq 0} \frac{ \sum_{k \in \mathcal{K}}\sum_{n \in \mathcal{S}_k} r_{k,n} }{\zeta P_T + P_C}(\alpha+(1-\alpha)\frac{\eta_P}{\eta_W})}
\end{equation}
\noindent \emph{subject to}
\begin{equation}\label{Pconstraint1}
{\sum_{n \in \mathcal{N}} \sum_{k \in \mathcal{S}_k} {r}_{k,n} \geq C_{0}}
\end{equation}
\begin{equation}\label{Pconstraint3}
{\sum_{k \in \mathcal{K}}\sum_{n \in \mathcal{S}_k}p_{k,n} = P_T \leq P_{max}  }
\end{equation}
\emph{has the following properties:}\\
\emph{(i) ${\lambda}_{RE}(P_T)$ is a continuously differentiable quasi-concave function with respect to $P_T$,}\\
\emph{(ii) the derivative of ${\lambda}_{RE}(P_T)$ meets the following condition
\begin{equation}\label{add1}
{\frac{d{\lambda}_{RE}(P_T)}{dP_T} = \frac{(\alpha+(1-\alpha)\frac{\eta_P}{\eta_W})\frac{d\bar{R}(P_T)}{P_T}-\alpha \zeta {\lambda}_{EE}(P_T)}{\zeta P_T + P_C}}
\end{equation}
where
\begin{equation}\label{add22}
{{\lambda}_{EE}(P_T) = \frac{\bar{R}(P_T)}{\zeta P_T + P_C},}
\end{equation}
and
\begin{equation}\label{add2}
{\bar{R}(P_T)\triangleq \max_{p_{k,n} \geq 0 } R(P_T) = \max_{p_{k,n}\geq 0 } \sum_{k \in \mathcal{K}}\sum_{n \in \mathcal{S}_k} r_{k,n}}
\end{equation}
represents the maximum sum rate under the maximum power constraint and minimum rate constraint (\ref{Pconstraint1})-(\ref{Pconstraint3}) meeting the condition
\begin{equation}\label{maximum sum rate derivative}
{\frac{d\bar{R}(P_T)}{P_T} = \max_{k \in \mathcal{K}, n \in \mathcal{S}_k} \frac{W_C{g}_{k,n}\log_2e }{1+ {p}^*_{k,n}{g}_{k,n}}}
\end{equation}
where ${g}_{k,n} \triangleq \frac{|h_{k}^n|^2}{\sigma^{n}_{k}} $ and ${p}^*_{k,n} (n \in \mathcal{S}_k)$ are respectively representing the channel-power-to-noise ratio (CNR) of the $k$-th UE on the $n$-th OFDMA subcarrier and the optimal allocated power on the $n$-th subcarrier to obtain $\bar{R}(P_T)$}. \\
\emph{Proof:} See Appendix C.

For the case with fixed transmission power $P_T$ and subcarrier assignment vector ${\boldsymbol{\rho}}$, we can rewrite the RE of the two-tier HCN as
\begin{equation}\label{rewrite}
{{\lambda}_{RE}(P_T) = \max_{p_{k,n}\geq 0} R(\frac{\alpha}{P} + (1-\alpha)\frac{W_{tot}}{W P_{tot}}) = \omega \bar{R}(P_T),}
\end{equation}
where $\omega \triangleq  \frac{\alpha}{P} + (1-\alpha)\frac{W_{tot}}{W P_{tot}}$. To derive the optimal power allocation, we can extend the multi-level water-filling scheme \cite{TangRE} to a HCN scenario as follows
\begin{equation}\label{water filling 1}
{\tilde{p}_{k,n} = (\mu_k - \frac{1}{g_{k,n}})^{+}, ~ \forall n \in \mathcal{S}_k ,}
\end{equation}
\begin{equation}\label{water filling 2}
{\sum_{n \in \mathcal{S}_k, \tilde{p}_{k,n} > 0} \sum_{k \in \mathcal{K}} W_C \log_2(\mu_k g_{k,n}) = C_0,}
\end{equation}
\begin{equation}\label{water filling 3}
{{p}^*_{k,n} = \tilde{p}_{k,n} + (\mu - \frac{1}{g_{k,n}} - \tilde{p}_{k,n})^{+},}
\end{equation}
\begin{equation}\label{water filling 4}
{\sum_{k \in \mathcal{K}} \sum_{{n \in \{\mathcal{S}_k | \bar{p}_{k,n} > \tilde{p}_{k,n}}\}} (\mu - \frac{1}{g_{k,n}} - \tilde{p}_{k,n}) = P_T - \sum_{k \in \mathcal{K}}\sum_{n \in \mathcal{S}_k} \tilde{p}_{k,n},}
\end{equation}
where $\mu_k$ and $\mu$ are used to denote the intermediate variables. The multi-level water-filling approach consists of two steps. Firstly, the power is allocated in order to satisfy the minimum rate requirement of the macro-cell UEs, where the allocated power in this step is $P_S = \sum_{k \in \mathcal{K}}\sum_{n \in \mathcal{S}_k} \tilde{p}_{k,n}$. Next, the remaining power is allocated in order to further improve the sum rate. Since quasi-concave function guarantees the existence of a unique maximum, we thus apply the gradient method to search for the optimal power. In particular, for a fixed subcarrier assignment set, gradient-based power adaptation can be used with single-UE water-filling in (\ref{water filling 1})-(\ref{water filling 2}), and the multi-level water-filling in (\ref{water filling 3})-(\ref{water filling 4}), where the power is updated using the gradient of RE
\begin{equation}\label{update power}
{P_T(n) = P_T(n-1) + t \times \frac{d{\lambda}_{RE}(P_T)}{dP_T}}
\end{equation}
with $t$ being the step size. Since the quasi-concavity property implies that ${\lambda}_{RE}(P_T)$ either strictly decreases or first increases and then strictly decreases with $P_T$, the proposed algorithm will terminate with either convergence or $P_0 = C_0^{-1}(\mathcal{S}_k, \gamma_k)$ if ${\lambda}_{RE}(P_T)$ is monotonically decreasing in $[P_0,P_{max}]$ and $P_{max}$ if ${\lambda}_{RE}(P_T)$ is monotonically increasing in $[P_0,P_{max}]$.

We are now ready to investigate the subcarrier assignment strategy. \emph{Theorem III} in Section IV can be applied here. Specifically, we assign each subcarrier to the MUE that would achieve the highest SINR on that subcarrier.
\begin{equation}\label{rho_alocation2}
\rho_{k}^{n*}~=~
\begin{cases}
1,~\textmd{if}~k = \textmd{arg}~\textmd{max}_{k \in \mathcal{K}}~  \gamma_{k}^n\\
0,~\textmd{otherwise}
\end{cases}.
\end{equation}
Note that the proposed power allocation approach only aims to solve the RE optimization problem (\ref{objectivepmre})-(\ref{constraint5pmre}) with fixed normalizing factor $\alpha$. Let ${\lambda}_{RE}(\alpha)$ denote the objective value of problem (\ref{objectivepmre})-(\ref{constraint5pmre}) with a given $\alpha$. It is easy to see that ${\lambda}_{RE}(1)$ denotes the maximum RE value of problem (\ref{objectivepmre})-(\ref{constraint5pmre}) that only aims to maximize EE without taking SE into account. On the other hand, ${\lambda}_{RE}(0)$ corresponds to the case in which SE is maximized without taking EE into account. Therefore, based on this result, we further develop a bi-section approach to numerically search for the optimal value of $\alpha$ per described in Table II.
\begin{table}
\centering
\caption{The proposed bi-section-based subcarrier assignment and power allocation approach.}
\renewcommand{\arraystretch}{1}  
\begin{tabular} {|l|}
\hline
1) ~Initialize $\alpha_{\textrm{min}} = 0$ and $\alpha_{\textrm{max}} = 1$;\\
2) ~~~\textbf{REPEAT};\\
3) ~~~Let $\alpha_{\textrm{mid}} = \frac{\alpha_{\textrm{min}}+\alpha_{\textrm{max}}}{2}$, perform the subcarrier assignment approach using (\ref{rho_alocation2});\\
4) ~~~~~\textbf{FOR} each UE $k \in K$ \\
5) ~~~~~Perform single-UE water-filling using (\ref{water filling 1})-(\ref{water filling 2}) to obtain $\tilde{p}_{k,n}$ and $\mu_k$, \\~~~~~~~~~calculate the power consumption $P_S$; \\
6) ~~~~~\textbf{END FOR} \\
7) ~~~~~\textbf{IF} $P_S > P_{max}$\\
8) ~~~~~~~$\textmd{infeasible}$; \\
9) ~~~~~\textbf{ELSE}   \\
10)~~~~~~~~Initial Power $P_T(1) \in [{P}_S, P_{max}]$;\\
11)~~~~~~\textbf{REPEAT}\\
12) ~~~~~~~~For the remaining power, perform the multi-level water-filling in (\ref{water filling 3})-(\ref{water filling 4});\\
13)~~~~~~~~~The transmission power is updated using the gradient of RE in (\ref{update power});\\
14)~~~~~~\textbf{UNTIL} when $|P_T(n)-P_T(n-1)| \leq \varepsilon$;  \\
15)~~~~\textbf{END IF}\\
16) ~\textbf{IF} $\lambda_{RE}(\alpha_{\textrm{mid}}) \geq \lambda_{RE}(\alpha_{\textrm{max}})$, let $\alpha_{\textrm{max}} = \alpha_{\textrm{mid}}$, ~\textbf{OTHERWISE} $\alpha_{\textrm{min}} = \alpha_{\textrm{mid}}$;\\
17) ~\textbf{UNTIL} $|\alpha_{max} - \alpha_{min}| \leq \varepsilon$. \\ \hline
\end{tabular} 
\end{table}
By employing the proposed bi-section based subcarrier assignment and power allocation approach, the optimal RE can be obtained for a given bandwidth (given number of subcarriers). As a result, starting from $W_{min} = KW_C$ (i.e., each UE should be guaranteed at least one subcarrier), we apply the proposed gradient-based power adaptation and the subcarrier allocation policy to the current bandwidth setting, and store the optimal RE value in the buffer ${\lambda}_{RE}(W)$. Then, we increment the bandwidth using $W = W + W_C$. The proposed subcarrier assignment and power allocation approach is performed again to obtain the maximum RE of the macro-cell under the updated bandwidth. This procedure is repeated for all possible bandwidth options, i.e., from $W_{min}$ to $W_{tot}$. Hence, the optimal RE and the corresponding bandwidth of the macro-cell is determined. The remaining bandwidth is then dedicated to the small-cell operation. A complete description of the algorithm can be found in Table III.
\begin{table}\centering
\caption{The proposed resource efficiency maximization scheme.} 
\renewcommand{\arraystretch}{1}  
\begin{tabular} {|l|}
\hline
1) ~\textbf{FOR} $W_c = W_{min} : W_{tot}$\\
2) ~~~Perform the proposed bi-section based subcarrier assignment\\~~~~~~ and power allocation approach;\\
3) ~~~Save the current RE value in the buffer;\\
4) ~\textbf{END FOR} \\
5) ~The maximum RE value and its corresponding bandwidth is determined. \\ \hline
\end{tabular} 
\end{table}

\subsection{Energy-Efficiency Optimization for Small-Cells}

Once the RE of the macro-cell is maximized, the remaining bandwidth is allocated exclusively to the small-cells. Since the macro-cell and small-cells are not sharing the same spectrum, there is no inter-tier interference. Moreover, considering the maximum transmit power of the small-cells is usually low (small coverage), and small-cells are geographically separated, the intra-tier interference between small-cells is considerably small. Therefore, considering the intra-tier interference as noise, we propose a low-complexity suboptimal resource allocation approach to maximize the EE of the small-cells.

Under this overlay-based setup, intra-tier interference is suppressed. The SINR expression can therefore be rewritten as
\begin{equation}
{\label{received_SINR_sub} \gamma_{[k,l]}^n = \frac{h_{[k,l,l]}^{n} p_{l}^{n}}{\sigma^{n}_{[k,l]}}}.
\end{equation}
As a result, the optimization problem in (\ref{objective})-(\ref{constraint4}) can be decomposed to a series of relatively isolated and simple optimization problems. In other words, one only needs to solve the EE maximization problem for each small-cell $(l \in \mathcal{L})$. This can be formulated as
\begin{align}
& \max_{\rho_{[k,l]}^{n}, p_{l}^{n}}~~ \frac{C_l}{\zeta P_T^{[l]} + P_C^{[l]}} \label{objectivesdd}\\
& \textrm{s.t.} ~~\sum_{n \in \mathcal{N}} \sum_{k \in \mathcal{K}_l} {p}_{l}^n\leq P_{max}^{[l]}, \label{constraint1sdd}\\
& ~~~~~~ \sum_{n \in \mathcal{N}} \sum_{k \in \mathcal{K}_l} \rho_{[k,l]}^{n}r_{[k,l]}^n \geq \delta_{small}, \label{constraint2sdd}\\
& ~~~~~~ \sum_{k \in \mathcal{K}_l} \rho_{[k,l]}^{n} = 1, \forall~n \in \mathcal{N} \label{constraint4sdd}
\end{align}
\textbf{\emph{Theorem V.}} \emph{The EE maximization problem in (\ref{objectivesdd})-(\ref{constraint4sdd}) is a special case of the RE maximization problem in (\ref{objectivepmre})-(\ref{constraint5pmre}).}\\
\emph{Proof:} With a given total transmit power, $P_T$, the maximum EE can be written as $\lambda_{EE} \triangleq  \frac{1}{(\alpha+(1-\alpha)\frac{\eta_P}{\eta_W})}\lambda_{RE}$. Since both $P_T$ and $W$ are constant, the maximum EE can be rewritten as $\lambda_{EE} = \nu\lambda_{RE}$. Therefore, the EE maximization problem in (\ref{objectivesdd})-(\ref{constraint4sdd}) is a special case of the RE maximization problem in (\ref{objectivepmre})-(\ref{constraint5pmre}). The proposed gradient-based power adaptation and the subcarrier allocation policy can be applied here to solve the EE optimization problem. \hspace{\fill} $\blacksquare$

As a result, we apply the proposed subcarrier allocation policy in (\ref{rho_alocation2}) for the remaining bandwidth, and then perform the multi-level water-filling algorithm in (\ref{water filling 1})-(\ref{water filling 4}) to obtain the optimal value. The power is then updated using the gradient of the EE as
\begin{equation}\label{update power11}
{P_T(n) = P_T(n-1) + t \times \frac{d{\lambda}_{EE}(P_T)}{dP_T}}
\end{equation}
where $t$ is the step size. Since ${\lambda}_{EE}(P_T)$ either strictly decreases or first increases and then strictly decreases with $P_T$, the proposed approach will terminate with either convergence or $P_0 = C_l^{-1}(\mathcal{S}_k, \gamma_k)$ if ${\lambda}_{EE}(P_T)$ is monotonically decreasing in $[P_0,P_{max}]$ and $P_{max}$ if ${\lambda}_{EE}(P_T)$ is monotonically increasing in $[P_0,P_{max}]$. This procedure is repeated for all the small-cells.

\section{Simulation Results}
\begin{table}[htp]
\centering
\caption{List of simulation parameters.}
\renewcommand{\arraystretch}{1}  
\centering
\begin{tabular}{|c|c|}
\hline
 Subcarrier bandwidth, $W_C$ & 15 KHz \\
 \hline
 Number of small-cells, L & 10 \\
 \hline
 Number of UEs at the macro-cell, $K_0$ & {10} \\
  \hline
 Number of UEs at the small-cells, $K_l$ & {3} \\
 \hline
 Maximum transmit power of macro-cell, $P_{max}^{[0]}$ & {46 dBm} \\
 \hline
 Maximum transmit power of small-cell, $P_{max}^{[l]}$ & {30 dBm} \\
 \hline
  Path loss from macro-BS to UEs & {$128.1 + 37.6\log_{10}d_M$ (dB) \cite{3GPPnew}} \\
 \hline
 Path loss from small-BS to UEs & {$140.7 + 36.7\log_{10}d_P$ (dB) \cite{3GPPnew}} \\
 \hline
\end{tabular}
\end{table}

\begin{figure}\centering
 \includegraphics{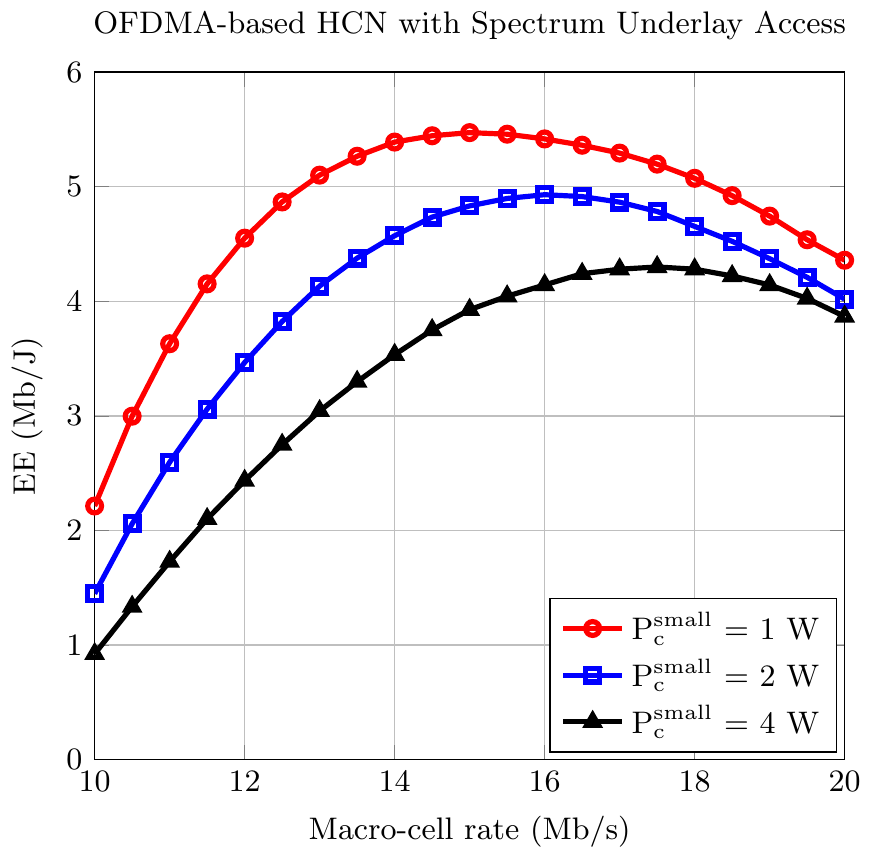}
 \caption{EE versus SE curves with different small-cell circuit power parameters.}
 \label{quasi-concave} 
\end{figure}

In this section, we present numerical results in order to verify our theoretical findings and analyze the performance of the proposed underlay and overlay approaches in terms of EE. It is assumed that ten uniformly-distributed small-cells are in the coverage area of a existing macro-cell, where ten and three uniformly-distributed UEs are serviced in the macro-cell and each small-cell, respectively. The radius of the macro-cell is set to 250 m, and that of the small-cells is set to 50 m. It should be noted that all results are obtained from various random locations of the UEs with identical and independent Rayleigh fading channels. The minimum throughput requirements for macro-cell and small-cells are set to 100 Mbps. Other simulation parameters are detailed in Table IV. In addition, these system parameters are merely chosen to demonstrate the EE optimization in an example and can easily be modified to any other values to address different scenarios.

\begin{figure}\centering
 \includegraphics{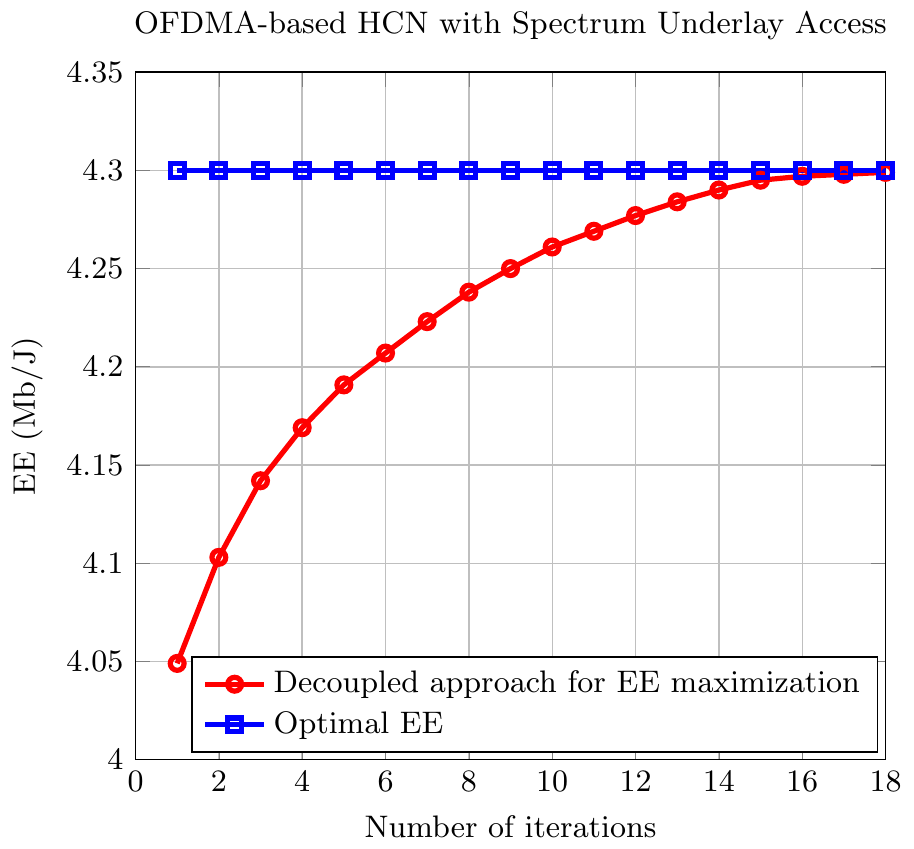}
 \caption{The performance of the proposed dual-layer approach with respect to optimal EE.}
 \label{convergence}
\end{figure}
\begin{figure}\centering
 \includegraphics{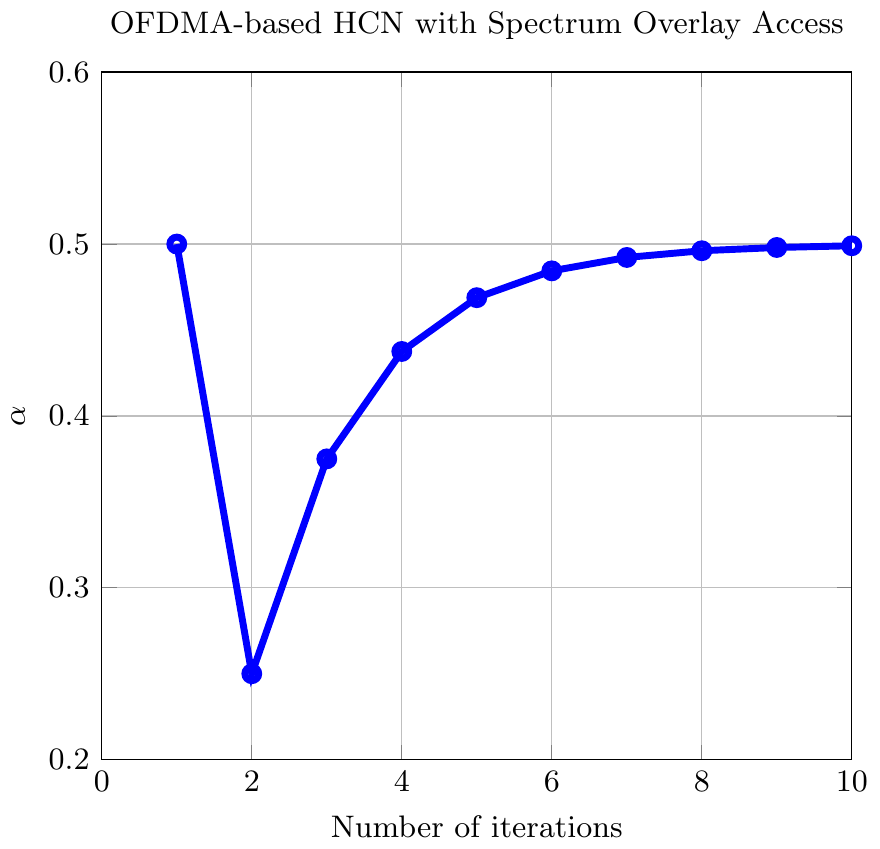}
 \caption{Convergence analysis of the proposed bi-section based approach.}
 \label{alpha} 
\end{figure}

In the first simulation, the performance of the proposed two solutions are studied. The EE-SE relationship (Theorem II) is first evaluated with different circuit powers (1, 2, 4 W) at the small-cells. It can be seen from Fig. \ref{quasi-concave} that the EE-rate relationship is quasi-concave and formes as a bell shape curve, where this quasi-concavity property is the basis of the proposed underlay-based approach. Furthermore, Fig. \ref{quasi-concave} also investigate the impact of circuit power on the EE-rate relationship. As anticipated, with increased circuit power, the corresponding optimal EE decreases due to higher power consumption. The performance of the proposed decomposition approach is then compared to the optimal EE. As it can be seen from Fig. \ref{convergence}, the proposed scheme successfully reaches the optimal EE after approximately 18 iterations. The validity of the proposed methodology is hence confirmed. On the other hand, Fig. \ref{alpha} depicts the convergence behavior of the proposed overlay-based approach. It is observed that the optimal $\alpha$ in this case is very close to 0.5, this is inline with our original work on RE in \cite{TangRE} where the amount of reduction is only $4\%$ on EE and $2\%$ on SE for an equal weight.

\begin{figure}\centering
 \includegraphics{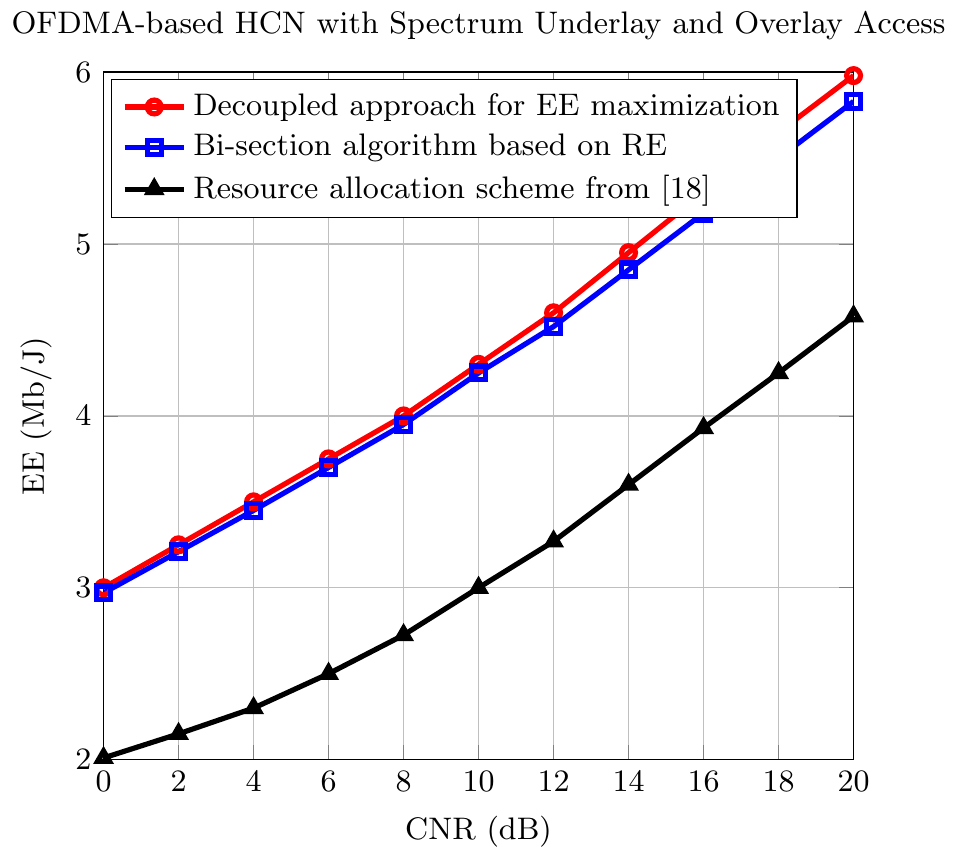}
 \caption{Comparison of different subcarrier assignment and power allocation schemes in terms of EE.}
 \label{comparison_CNR} 
\end{figure}

Next, we evaluate the EE performance of the different schemes with spectrum underlay and overlay access. For comparison purposes, we pair the proposed schemes against the joint subcarrier assignment and power allocation scheme in \cite{DuyTrongNgo}. The optimal EE is evaluated across the 0-20 dB CNR range. As it can be seen from Fig. \ref{comparison_CNR}, the EE achieved by the proposed bisection-based approach is very close to that of the proposed decomposition resource allocation approach whilst being much more effcient in terms of computational complexity. It is importan to highlight, however, that the performance gap increases in high CNR region. This is because the overlay-based approach treats the intra-tier interference as noise, and will become dominant when the noise power diminishes at high CNR regime; hence resulting in reduced EE performance. Furthermore, both algorithms achieve higher EE compared to the scheme proposed in \cite{DuyTrongNgo} which aims to maximize the sum rate.

\begin{figure}\centering
\includegraphics{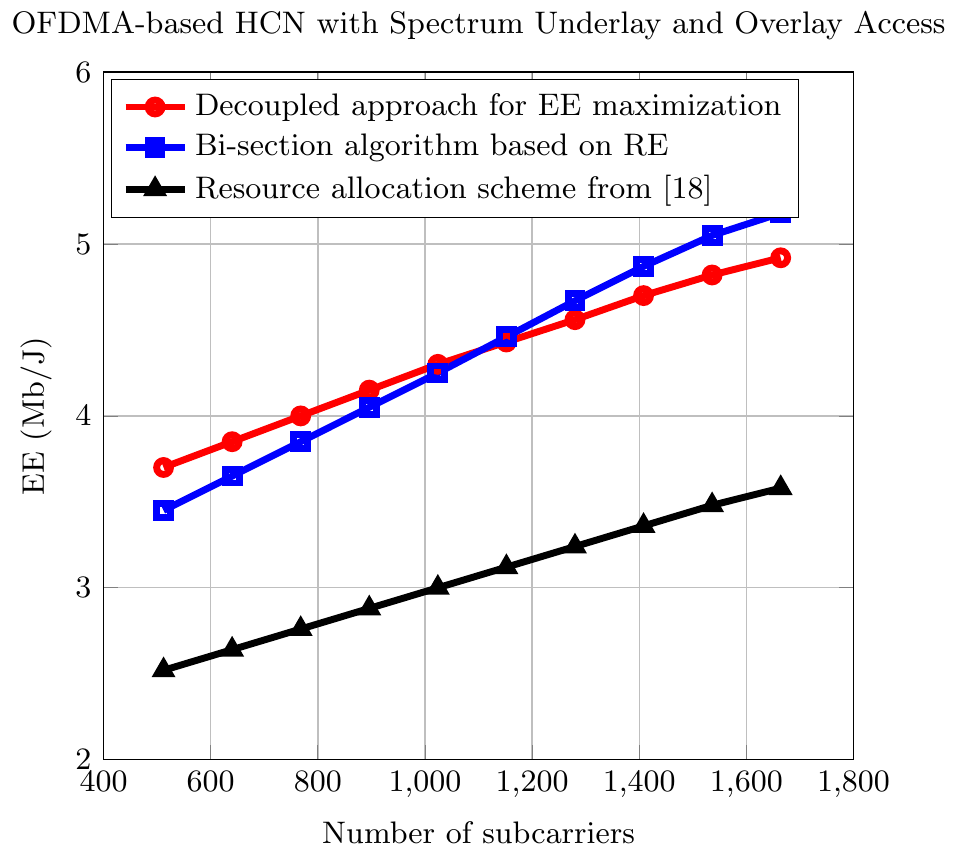}
\caption{Impact of the number of subcarriers on the EE performance of different schemes.}
\label{comparison_subcarriers} 
\end{figure}

The impact of the number of subcarriers on the optimal EE is illustrated in Fig. \ref{comparison_subcarriers}. It can be seen that with a moderate number of subcarriers (small bandwidth), e.g., $N \leq 1000$, the EE achieved by the proposed overlay-based approach is lower than that of the proposed approach using underlay transmission. However this trend is reversed when the system has a larger bandwidth (e.g., $N \geq 1000$). The reason for this observation is that the proposed underlay-based approach allocates all available bandwidth to all cells. The excessive transmission-associated circuit power, which is modeled as a linear function of the bandwidth, will reduce the EE performance in a system with large bandwidth. On the other hand, the proposed overlay-based approach allocates the exclusive spectrum parts to the macro-cell and small-cells and hence is more suitable for in the context of bandwidth-abundant HCNs.

\begin{figure}\centering
 \includegraphics{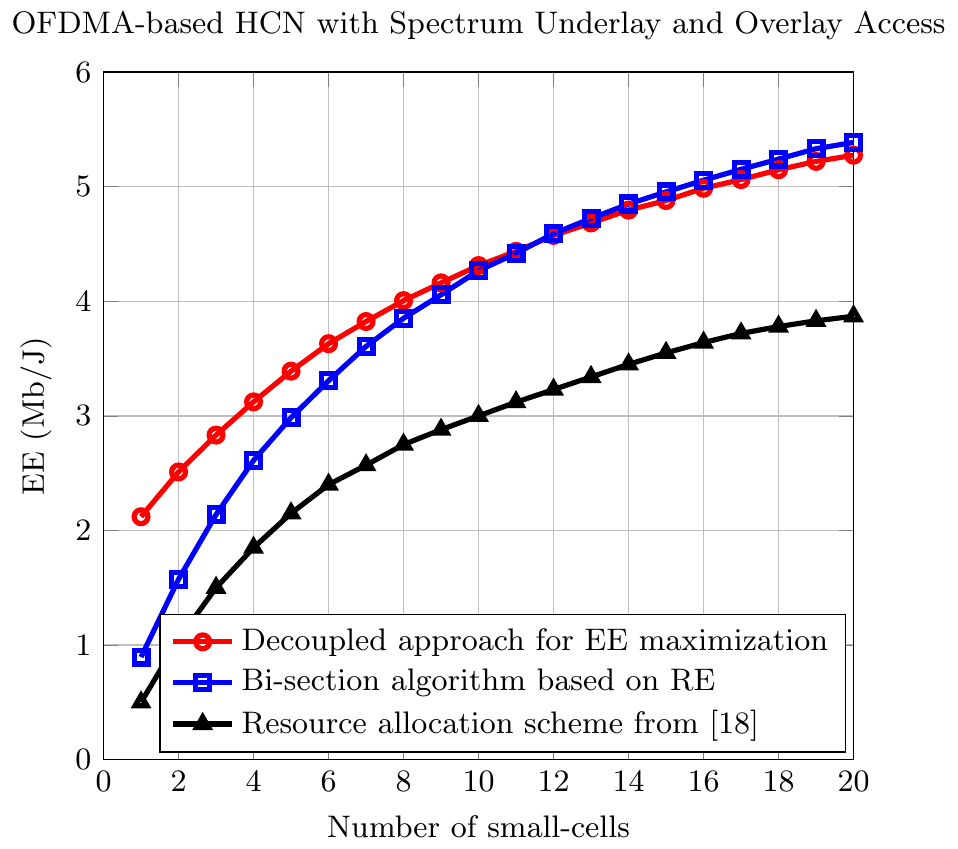}
 \caption{Impact of small-cells density on the EE performance of different schemes.}
 \label{comparison_cell} 
\end{figure}

Finally, the impact of small-cells density on the optimal EE is investigated in Fig. \ref{comparison_cell}. As shown in the figure, with a lower small-cells density, e.g., $L \leq 10$, the EE achieved by the proposed bisection-based approach is lower than that of the proposed decomposed resource allocation approach. However, for the case of dense small-cells, e.g., $L \geq 10$, the EE achieved by the strategy using RE and overlay transmission is superior. This is because when the small-cells density is low, the inter-tier interference will have less impact resulting in lower transmit power levels needed to satisfy the QoS targets. Therefore, the underlay transmission strategy is suitable under this setup. On the other hand, for the case of a dense deployed HCN, e.g., with $L \geq 10$, the inter-tier interference will become significant when spectrum is shared by the different tiers. Hence, extra power is required to maintain the throughput requirements of the UEs. Moreover, since macro-cell and small-cells occupy the whole bandwidth at the same time, the excessive transmission associated circuit power will further degrade EE performance. Consequently, the proposed overlay-based approach where different portions of the spectrum is allocated exclusively to the macro-cell and the small-cells is more suitable for dense multi-tier cellular environments.

\section{Conclusions}

In this paper, we have addressed the EE optimization problem for OFDMA-based two-tier HCNs consisting of a macro-cell and multiple small-cells. Subcarrier assignment and power allocation policies were jointly investigated to optimize EE considering spectrum underlay and overlay access. Considering underlay transmission, where macro-cell and small-cells are sharing the available spectrum, we proved the relationship between EE and achievable rate is a quasi-concave function. On the basis of this property, we decomposed the original problem with multiple inequality constraints into multiple optimization problems with single inequality constraints. For each sub-problem, we separated the subcarrier assignment and power allocation process and developed an optimal solution based on difference-of-two-concave-functions approximation, successive convex approximation, and gradient-search methods. On the other hand, the underlay approach may not be energy-efficient due to severe inter-tier interference in a dense HCN scenario. In addition, it will lead to a higher power consumption in a bandwidth-abundant system and hence reduce the EE performance. Therefore, we developed a novel low-complexity resource allocation scheme based on the idea of overlay transmission and RE. In this approach, we first optimized the RE of macro-cell and determined the optimal corresponding bandwidth, we then allocated the remaining bandwidth to small-cells and optimized the EE. Simulation results confirmed the theoretical findings and demonstrated that the proposed algorithms can efficiently approach the optimal EE.

\section*{Appendix A}
\begin{center}
    \textsc{Proof of Theorem I}
 \end{center}
To prove ${\lambda}^*_{EE}(\textbf{C})$ is a quasi-concave function, we denote the superlevel sets of ${\lambda}^*_{EE}(\textbf{C})$ as
\begin{equation}\label{quasi-concave_example}
{\mathcal{S}_\kappa = \{\textbf{C} \geq \boldsymbol{\delta} | {\lambda}^*_{EE}(\textbf{C}) \geq \kappa \}.}
\end{equation}
In accordance with \cite{Boyd04}, for any real number $\kappa$, if the convexity for $\mathcal{S}_\kappa$ is satisfied, ${\lambda}^*_{EE}(\textbf{C})$ is strictly quasi-concave in $\textbf{C}$. Therefore, we here divide the proof into two cases. For the case of $\kappa< 0$, since EE is always positive and hence there are no points on the counter, ${\lambda}^*_{EE}(\textbf{C}) = \kappa$. For the case of $\kappa \geq 0$, $\lambda_{EE}$ can be rewritten as
\begin{equation}\label{EE_proof}
{\lambda_{EE} = \frac{\sum_{l \in \mathcal{L}} C_l }{\zeta P_T(\textbf{C}) + P_C},}
\end{equation}
and hence $\mathcal{S}_\kappa$ can be rewritten as $\kappa \zeta  P_T(\textbf{C}) + \kappa P_C - \sum_{l \in \mathcal{L}}C_l$ $\leq 0$. In \cite{TangJSAC}, its been proved that $P_T(\textbf{C})$ is convex in $\textbf{C}$, therefore the convexity property of $\mathcal{S}_\kappa$ holds and ${\lambda}^*_{EE}(\textbf{C})$ is strictly quasi-concave in $\textbf{C}$. This completes the proof of \emph{Theorem I}.\hspace{\fill} $\blacksquare$

\section*{Appendix B}
\begin{center}
    \textsc{Proof of Theorem III}
 \end{center}
Suppose that when we obtain the optimal solution of problem in (\ref{objectivepms})-(\ref{constraint3pms}), where for cell $l \in \mathcal{L}$, subcarrier $n \in \mathcal{N}$ is allocated to UE $k \in \mathcal{K}_l\setminus \{k^*(n,l)\}$, where $k^*(n,l)$ represents the UE that consumes the lowest power on subcarrier $n$. However, to maintain the minimum rate demand, if $n$ is instead allocated to $k^*(n,l)$ and $p^n_{k^*(n,l),l} < p^n_{k,l}$, the interference received by the UEs that use $n$ will be decreased, hence reducing the power consumption for all UEs. This statement contradicts the initial assumption that the optimal assignment for $n \in \mathcal{N}$ is allocated to UE $K \in \mathcal{K}_l\setminus \{k^*(n,l)\}$. Therefore, subcarrier $n$ should be assigned to $k^*(n,l)$. This completes the proof of \emph{Theorem III}. \hspace{\fill} $\blacksquare$

\section*{Appendix C}
\begin{center}
    \textsc{Proof of Theorem IV}
 \end{center}
With a given subcarrier assignment set ${\boldsymbol{\rho}}$, ${\bar{\lambda}_{RE}(P_T)}$ could be rewritten as $\alpha\frac{\bar{R}(P_T)}{P} + \nu \bar{R}(P_T)$, where $\nu = (1-\alpha)\frac{W_{tot}}{W P_{tot}}$. Under the water-filling approach, the transmit power on each subcarrier is non-decreasing. With the assumption that $\sum_{k \in \mathcal{K}} \sum_{n \in \mathcal{S}_k}\triangle p_{k,n} = \triangle P_T$, the existence of the limit reveals that $\bar{R}(P_T)$ is continuously differentiable and satisfies the following equation
\begin{equation}\label{rate gradient}
\frac{d\bar{R}(P_T)}{dP_T} = \frac{d\bar{R}(P_T)}{dp_{k,n}} = \max_{k \in \mathcal{K}, n \in \mathcal{S}_k} \frac{W_C g_{k,n} \log_2 e}{1+g_{k,n}\bar{p}_{k,n}}.
\end{equation}
Furthermore, for the case of $k \in \mathcal{K}$ and $n \in \mathcal{S}_k$, $\frac{W_C g_{k,n} \log_2 e}{1+g_{k,n}\bar{p}_{k,n}}$ is non-increasing with $P_T$ whilst $\max_{k \in \mathcal{K}, n \in \mathcal{S}_k} $ $\frac{W_C g_{k,n} \log_2 e}{1+g_{k,n}\bar{p}_{k,n}}$ is decreasing with respect to $P_T$. Hence, we can conclude that $\frac{d^2\bar{R}(P_T)}{dP_T^2} < 0$ and $\bar{R}(P_T)$ is a strictly concave function with $P_T$.

Similar to the proof in Appendix A, if the convexity for $\mathcal{S}_\theta$ holds, $\frac{\bar{R}(P_T)}{P_T}$ is strictly quasi-concave in $P_T$. Therefore, we here divide the proof into two cases. For the case of $\theta< 0$, since rate is always positive and hence there are no points on the counter, $\frac{\bar{R}(P_T)}{P_T} = \theta$. For the case of $\theta \geq 0$, $\mathcal{S}_\theta$ can be rewritten as $\mathcal{S}_\theta = \{P_T \geq \sum_{k \in \mathcal{K}} R^{-1}(\mathcal{S}_k, C_l) | \theta \zeta P_T + \theta P_C - \bar{R}(P_T)\} \leq 0$, where $R^{-1}(\mathcal{S}_k, C_l)$ denotes the minimum transmit power required to satisfy the throughput demand $C_l$. Therefore, given that the concavity of $\bar{R}(P_T)$ holds and $\mathcal{S}_\theta$ is strictly convex in $P_T$, $\frac{\bar{R}(P_T)}{P_T}$ is continuously differentiable and quasi-concave with respect to $P_T$, this concludes the proof of Property (i) in \emph{Theorem IV}.

Moreover, considering ${\bar{\lambda}_{RE}(P_T)}$ as $\alpha\frac{\bar{R}(P_T)}{P} + \nu \bar{R}(P_T)$, the derivative of RE $\frac{d\bar{\lambda}_{RE}(P_T)}{dP_T}$ should satisfy the following equation
\begin{equation}\label{dEE331}
{\frac{d\bar{\lambda}_{RE}(P_T)}{dP_T} = \alpha\frac{d\frac{\bar{R}(P_T)}{P}} {dP_T}+\nu \frac{d\bar{R}(P_T)}{dP_T}}.
\end{equation}
Thus, based on (\ref{dEE331}), $\frac{d\frac{\bar{R}(P_T)}{P}} {dP_T}$ can be further constructed as follows
\begin{align}
\frac{d\frac{\bar{R}(P_T)}{P}} {dP_T} & = \lim_{\triangle P_T \rightarrow 0} \frac{ \frac{\bar{R}(P_T+\triangle P_T)}{\zeta (P_T+\triangle P_T) + P_C} -  \frac{\bar{R}P_T}{\zeta P_T + P_C} }{\triangle P_T} = \lim_{\triangle P_T \rightarrow 0} \frac{\frac{\bar{R}(P_T+\triangle P_T)-\bar{R}P_T}{\triangle P_T} - \zeta \bar{\lambda}_{EE(P_T)}}{\zeta (P_T+\triangle P_T) + P_C}  \label{dEE1} \\
& = \frac{ \frac{d\bar{R}(P_T)}{dP_T}- \zeta \bar{\lambda}_{EE}(P_T)}{\zeta P_T + P_C} . \label{dEE3}
\end{align}
Hence, we observe
\begin{equation}\label{dRE}
{\frac{d\bar{\lambda}_{RE}(P_T)}{dP_T} = \frac{(\alpha+(1-\alpha)\frac{\eta_P}{\eta_W})\frac{d\bar{R}(P_T)}{P_T}-\alpha \zeta \bar{\lambda}_{EE}(P_T)}{\zeta P_T + P_C}}
\end{equation}
\noindent where $\frac{d\bar{R}(P_T)}{dP_T} = \max_{k \in \mathcal{K}, n \in \mathcal{S}_k} \frac{W_C g_{k,n} \log_2 e}{1+g_{k,n}\bar{p}_{k,n}}$, $\bar{\lambda}_{EE}(P_T) = \frac{\bar{R}(P_T)}{\zeta P_T + P_C}$. This concludes the proof of Property (ii) in \emph{Theorem IV}.\hspace{\fill} $\blacksquare$

\bibliographystyle{IEEEtran}
\bibliography{IEEEabrv,references}

\end{document}